\documentclass[twocolumn,amstext,amssymb,superscriptaddress,floatfix,showpacs,nofootinbib]{revtex4-1}
\usepackage{amsmath}
\usepackage{lineno} 
\usepackage{ulem}

\usepackage{todonotes}
\presetkeys{todonotes}{color=green!40, size=\footnotesize}{}
\usepackage{hyperref}

\hypersetup{colorlinks=true,citecolor=blue,citebordercolor=red,linktoc=all,linkcolor=blue}

\usepackage{setspace}

\usepackage{amssymb,amsfonts,multirow}

\usepackage{xcolor,colortbl}

\definecolor{Gray}{gray}{0.85}
\definecolor{LightGreen}{rgb}{0.88, 1, 0.88}
\definecolor{Blue}{rgb}{0,1,1}
\definecolor{Lime}{rgb}{0,1,0}
\definecolor{LightCyan}{rgb}{0.88,1,1}
\definecolor{LightRed}{rgb}{1, 0.85, 0.85}
\definecolor{Red}{rgb}{1, 0, 0}
\definecolor{LightYellow}{rgb}{1, 1, 0.85}
\definecolor{Yellow}{rgb}{1,1,0.05}
\definecolor{LightBlue}{rgb}{0.87, 0.94, 1}
\definecolor{white}{gray}{1}
\definecolor{black}{gray}{0}

\definecolor{LightGray}{gray}{0.93}

\usepackage{array,mathtools,amssymb,booktabs}
\newcolumntype{C}{>{$}c<{$}}
\AtBeginDocument{
\heavyrulewidth=.16em
\lightrulewidth=.1em
\cmidrulewidth=.03em
\belowrulesep=.4ex
\belowbottomsep=0pt
\aboverulesep=.4ex
\abovetopsep=0pt
\cmidrulesep=\doublerulesep
\cmidrulekern=.5em
\defaultaddspace=.5em
}
\newcolumntype{G}{>{\columncolor{LightGray}}c}

\usepackage{amsmath}
\usepackage{amssymb}
\usepackage{amsfonts}
\usepackage{bm}
\usepackage{graphicx}

\usepackage{bbm}

\usepackage{slashed}

\newcommand\numberthis{\addtocounter{equation}{1}\tag{\theequation}}
\newcommand{\Tr}{\operatorname{Tr}}

\newcommand{\op}{\operatorname}

\newcommand{\hc}{\op{h.c.}}

\newcommand{\ybold}{\mathbf{Y}}

\makeatletter
    \def\CT@@do@color{%
      \global\let\CT@do@color\relax
            \@tempdima\wd\z@
            \advance\@tempdima\@tempdimb
            \advance\@tempdima\@tempdimc
    \advance\@tempdimb\tabcolsep
    \advance\@tempdimc\tabcolsep
    \advance\@tempdima2\tabcolsep
            \kern-\@tempdimb
            \leaders\vrule
                    \hskip\@tempdima\@plus  1fill
            \kern-\@tempdimc
            \hskip-\wd\z@ \@plus -1fill }
    \makeatother

\begin{document}
${}$\vskip1cm

\title{Price of Asymptotic Safety}
\author{Andrew~D.~Bond}
\email{a.bond@sussex.ac.uk}

\author{Daniel F.~Litim}
\email{d.litim@sussex.ac.uk}
\affiliation{\mbox{Department of Physics and Astronomy, U Sussex, Brighton, BN1 9QH, U.K.}}

\begin{abstract}
All known examples of four dimensional quantum field theories with asymptotic freedom or 
asymptotic safety at weak coupling  
involve non-abelian gauge interactions. We demonstrate that this is not a coincidence: no weakly coupled fixed points, ultraviolet or otherwise, can be reliably generated in theories lacking gauge interactions. Implications for  particle physics, critical phenomena, and conformal field theory, 
are  indicated.
\end{abstract}

\maketitle

{\it Introduction.---}
A  turning point in the understanding of   high-energy physics has been the discovery of asymptotic freedom in non-abelian gauge theories \cite{Gross:1973id, Politzer:1973fx}.  It ensures that certain renormalisable quantum field theories remain predictive  in the high-energy  limit  where couplings becomes  free \cite{Gross:1973ju,Cheng:1973nv,Chang:1974bv}. 
Non-abelian gauge fields are decisive for this to happen:
without them, asymptotic freedom cannot be achieved in any theory involving Dirac fermions, photons, or scalars \cite{Coleman:1973sx}.  

In the absence of asymptotic freedom, particle theories are  generically plagued by divergences  and a breakdown of predictivity  in the high-energy limit. 
Some such theories, however, remain well-defined thanks to strict cancellations at the quantum level  \cite{Litim:2014uca,Bond:2016dvk} and display ``asymptotic near freedom''  \cite{Bailin:1974bq}
or ``asymptotic safety'' \cite{Weinberg:1980gg} at high energies.
Thereby,  running couplings achieve  an {\it interacting} 
fixed point
under the renormalisation group evolution, 
which serves as an anchor for short distance quantum fluctuations  \cite{Wilson:1971bg}.
  General theorems for asymptotic safety
 are available for  weakly coupled gauge-matter theories 
    \cite{Bond:2016dvk} 
    and cover
simple \cite{Litim:2014uca,Bond:2017tbw}, semi-simple  \cite{Bond:2017lnq}, and supersymmetric  \cite{Bond:2017suy} gauge theories, and  extensions beyond the Standard Model  \cite{Bond:2017wut}.
For studies also involving quantum gravity, see \cite{Reuter:1996cp, Litim:2003vp,Niedermaier:2006ns,Litim:2011cp,Benedetti:2009gn, Falls:2013bv,Falls:2014tra,Falls:2016wsa,Falls:2017lst,Percacci:2017fkn,Robinson:2005fj,Shaposhnikov:2009pv,Niedermaier:2010zz,
Folkerts:2011jz,Dona:2013qba,Christiansen:2017cxa,Eichhorn:2018yfc} and references therein.

It appears that all known examples of four-dimensional particle theories with asymptotic freedom or  asymptotic safety at weak coupling involve non-abelian gauge interactions. It is the purpose of this  Letter to demonstrate that this is not a coincidence: no weakly interacting fixed points, ultraviolet or otherwise, can be reliably generated in theories lacking gauge interactions. 
Partial results in support of our claim have been made available in \cite{Coleman:1973sx,Bond:2016dvk}. Here, we provide the missing pieces which are, on the one  hand, an extension of the Coleman-Gross theorem \cite{Coleman:1973sx},
and  a no-go-theorem for weakly interacting fixed points in non-gauge theories, on the other.
Taken together, non-abelian gauge interactions are the unique price for particle theories to remain  strictly perturbative and predictive at asymptotically high energies, and to display weakly coupled fixed points at low energies.

{\it Price of asymptotic freedom.---}
To establish our claim, we first revisit asymptotic freedom of general, renormalisable particle theories in four dimensions involving gauge fields, fermions, or scalars. Without loss of generality, we  limit the analysis   to the canonically marginal interactions  which are
the gauge, the Yukawa, and the scalar self-couplings $\{g_i,{\bf Y}^A_{IJ},\lambda_{ABCD}\}$,  respectively.
 We assume canonically normalised kinetic terms with gauge couplings $g_i$ for each gauge factor.  Our conventions for the most general Yukawa and scalar couplings are
  \begin{equation}\label{Yukpot}
 \begin{array}{rcl}
  L_{\rm Yuk.}&=&-\tfrac{1}{2}(\ybold^A_{JK}\Phi^A\Psi_J\Psi_K + \hc)\,,\\[1ex]
  L_{\rm pot.}&=& -\frac{1}{4!}\lambda_{ABCD}\Phi^A\Phi^B\Phi^C\Phi^D\,,
  \end{array}
  \end{equation}
  where $\Psi_J$ denote Weyl fermions, and $\Phi^A$ real scalars. Matter fields may be charged under the gauge groups.

Next, we turn to quantum effects and the renormalisation group running of couplings. The point in coupling space where all couplings vanish, the free theory, is always a fixed point of the renormalisation group. Then, for any theory to be free at asymptotically high energies, the free fixed point must be ultraviolet and the beta functions negative for sufficiently small couplings,
\begin{align}\label{negative}
\mu\,\partial_\mu\left(g,{Y},\lambda\right)< 0\,,
\end{align}
with $\mu$ the renormalisation group scale. After scaling the  loop factor into the couplings, as we shall consistently do throughout,
the  one-loop gauge beta functions  is \cite{Gross:1973id, Politzer:1973fx,Gross:1973ju}
 \begin{equation}\label{gauge}
 \mu\,\partial_\mu g_i=B_i\,g_i^3\,,\quad B_i=-\tfrac{11}{3} C_2^{G_i} + \tfrac{2}{3} S_2^{F_i} + \tfrac{1}{6}S_2^{S_i}\,.
 \end{equation}
Non-abelian gauge fields contribute negatively to the one loop coefficient $(B_i)$, proportionally to  the 
Casimir of the gauge group 
$(C_2^{G_i})$. Matter fields contribute positively and proportionally to their Dynkin indices  $(S_2)$.
    The main feature of non-abelian theories is that \eqref{gauge} can have either sign. 
    A negative one loop coefficient 
    is known to offer the {\it unique} key for  asymptotic freedom  
    \cite{Gross:1973ju,Cheng:1973nv,Chang:1974bv}.
Below, we demonstrate that non-abelian  fluctuations, in particular the smallness of the one loop coefficient
for suitable matter
 and irrespective of the sign of \eqref{gauge},  also provide the {\it unique} key  for weakly interacting fixed points including asymptotic safety of theories 
 with $B_i< 0$.

{\it Coleman-Gross theorem revisited.---}
 To clarify the role of  gauge field fluctuations  for asymptotic freedom, 
 we first revisit the Coleman-Gross theorem \cite{Coleman:1973sx}. It states that a non-gauge theory of scalars with or without Dirac fermions cannot become asymptotically free. To cover the most general setting, we
 extend the theorem towards Weyl fermions. It is convenient to view the Yukawa couplings as symmetric matrices $\ybold^A$  in the fermion indices $(\ybold^A)_{JK}\equiv\ybold^A_{JK}$.
Their running with momentum scale $\mu$ at the leading order in perturbation theory reads \cite{Machacek:1983fi,Luo:2002ti}
\begin{widetext}
\begin{align}\label{betaY}
	\mu\,\partial_\mu \ybold^A &= \tfrac{1}{2}\left(\overline{\ybold_{\!\bm2}^{\!\bm F}}\,\ybold^A + \ybold^A\,\ybold_{\!\bm2}^{\!\bm F} \right)  + Y_2^{S\,AB}\,\ybold^B+ 2 \ybold^B\, \ybold^{A\dagger}\,\ybold^B\,,
\end{align}
where the bar denotes complex conjugation, and summation over repeated indices is implied. We have also introduced the quadratic combinations
$\ybold^{\,\,\,\,AB}_{\!\bm2\,JK} = \tfrac{1}{2}\left(\ybold^{A\dagger}\ybold^B + \ybold^{B\dagger}\ybold^A\right)_{JK}$ alongside $\ybold^{\!\bm F}_{\!\bm2\,JK} \equiv \ybold^{\,\,\,AA}_{\!\bm2\,JK}$ and
$Y_2^{S\,AB} \equiv \ybold^{\,\,\,AB}_{\!\bm2\,JJ}$.
The first and second  term in \eqref{betaY} arise from the wave function renormalisation of the fermion and scalar propagators, whereas the last term stems from vertex corrections. 
We note that the Yukawa couplings and their flows \eqref{betaY} transform as tensors under a change of base, $i.e.$~general linear transformations of the  fields which leave \eqref{Yukpot} invariant. 

We shall now focus our attention on the flow for the sum of the squared absolute values  of all Yukawa couplings, 
\begin{align}\label{eq:sumbeta}
	\mu\,\partial_\mu \Tr(\ybold^{A\dagger}\ybold^A)
		&=  
		\Tr[(\overline{\ybold_{\!\bm2}^{\!\bm F}})^2] + \Tr[(\ybold_{\!\bm2}^{\!\bm F})^2] 
			+ 4 \Tr(\ybold^{A\dagger}\ybold^B\ybold^{A\dagger}\ybold^B) 
			+
\Tr(\ybold^{A\dagger}\ybold^B)\left[\Tr(\ybold^{A\dagger}\ybold^B) +(A\leftrightarrow B)
\right]
\end{align}
for if we are to have all Yukawa beta functions negative, then  this combination must be negative as well.  We emphasize that the flow
\eqref{eq:sumbeta} 
 and the  conclusions  drawn from it are independent of the choice of field base.
A lower bound for \eqref{eq:sumbeta} follows by using that ${\rm Re} \, z^2 \le z^*z$ for any complex number $z$, whence
\begin{align}\label{zz}
	\Tr(\ybold^{A\dagger}\ybold^B)\Tr(\ybold^{A\dagger}\ybold^B) 
		\leq 
		 \Tr(\ybold^{A\dagger}\ybold^B)\Tr(\ybold^{B\dagger}\ybold^A)\,.
\end{align}
Next, we introduce the three real trace invariants 
$T_1=\Tr[(\ybold_{\bm 2}^{\!\bm F})^2]= \Tr[(\overline{\ybold_2^{\!\bm F}})^2]$, 
$T_2=\Tr(\ybold^{A\dagger}\ybold^B\ybold^{A\dagger}\ybold^B)$, and
$T_3=\Tr(\ybold^{A\dagger}\ybold^B)\Tr(\ybold^{A\dagger}\ybold^B)$. By definition, $T_2$ may have either sign while $T_1,T_3\ge 0$. In terms of these, and together with \eqref{zz}, 
we find that the  Yukawa beta function \eqref{eq:sumbeta} is bounded from below,
\begin{align}\label{eq:betabound}
	\mu\,\partial_\mu \Tr(\ybold^{A\dagger}\ybold^A)
		&\geq 2( T_1 + T_2) + 2(T_2+T_3)\,.
\end{align}
Recalling that the Yukawa couplings are symmetric in the fermionic indices, 
we rearrange the sums as follows
\begin{align*}
	T_1 + T_2 &= \ybold^{A\dagger}_{JK}\ybold^A_{KL}\ybold^{B\dagger}_{LM}\ybold^B_{MJ}
				+\ybold^{A\dagger}_{JK}\ybold^B_{KL}\ybold^{A\dagger}_{LM}\ybold^B_{MJ}
		= \ybold^{A\dagger}_{JK}\ybold^B_{MJ}
			\Big(\ybold^A_{KL}\ybold^{B\dagger}_{LM} + \ybold^B_{KL}\ybold^{A\dagger}_{LM}\Big)=
			\\ &
		= \tfrac{1}{2}\Big( \ybold^{A\dagger}_{JK}\ybold^B_{MJ} +  \ybold^{B\dagger}_{JK}\ybold^A_{MJ}\Big)
				\Big(\ybold^A_{KL}\ybold^{B\dagger}_{LM} + \ybold^B_{KL}\ybold^{A\dagger}_{LM}\Big)
		= 2 \ybold_{\!\bm2\,KM}^{\,\,\,\,AB} \ybold_{\!\bm2\,MK}^{\,\,\,\,AB} 
			= 2\overline{\ybold_{\!\bm2\,MK}^{\,\,\,\,AB}}\ybold_{\!\bm2\,MK}^{\,\,\,\,AB} 
			\,,\label{eq:T12}\numberthis \\[1ex]		
	T_2 + T_3 &=  \ybold^A_{JK}\ybold^{B\dagger}_{KL}\ybold^A_{LM}\ybold^{B\dagger}_{MJ}
	+\ybold^A_{JK}\ybold^{B\dagger}_{KJ}\ybold^A_{LM}\ybold^{B\dagger}_{ML}
		= \ybold^A_{JK}\ybold^A_{LM}\left(\ybold^{B\dagger}_{KJ}\ybold^{B\dagger}_{ML}
				+\ybold^{B\dagger}_{KL} \ybold^{B\dagger}_{MJ}\right)=\\
		&= \tfrac{1}{2}\Big(\ybold^A_{KJ}\ybold^A_{ML} + \ybold^A_{KL}\ybold^A_{MJ}\Big)
			\Big(\ybold^{B\dagger}_{KJ}\ybold^{B\dagger}_{ML}
				+\ybold^{B\dagger}_{KL} \ybold^{B\dagger}_{MJ}\Big) 
				\,.\numberthis
				\label{eq:T23}
				\end{align*}
\end{widetext}
As is evidenced by the explicit expressions, both \eqref{eq:T12} and \eqref{eq:T23} are sums of absolute values squared and 
therefore manifestly semi-positive definite,
\begin{equation}\label{eq:tracebounds}
	\begin{array}{rl}
	T_1 + T_2 &\geq 0\,, \\[1ex] T_2 + T_3 &\geq 0\,.
	\end{array}
\end{equation}
Most importantly, the bounds \eqref{eq:tracebounds}  dictate  positivity for the flow \eqref{eq:betabound} close to the Gaussian,
\begin{align}\label{bound}
	\mu\,\partial_\mu \Tr(\ybold^{A\dagger}\ybold^A)
		&
		\geq 0\,,
\end{align}
and establish that asymptotic freedom is unavailable.
Had we substituted Weyl by Dirac fermions in \eqref{Yukpot}, we would have found the lower bound $\mu\partial_\mu \Tr(\ybold^{A\dagger}\ybold^A)\geq  2T_1 + 4 (T_2 + T_3)$,  instead of  \eqref{eq:betabound}. 
For theories with Dirac fermions only, the non-negativity of $T_1$ together with $T_2+T_3>0$ is sufficient to conclude the absence of asymptotic freedom  \cite{Coleman:1973sx}. Clearly, the bounds for Weyl and Dirac fermions  are inequivalent: while the former entail the latter,  the converse is not true.

One might wonder whether  scalar self-interactions may upset the conclusion.
Scalar couplings contribute to the Yukawa beta function starting at two-loop order. Therefore, if they were to reliably generate asymptotic freedom, they must do so along a renormalisation group trajectory where they are parametrically larger than the Yukawa couplings. Assuming this to be the case,  we can then  ignore the Yukawa contribution to the running of the quartics. In other words,  the scalar sector must become asymptotically free in its own right. This, however, is known to be impossible  \cite{Coleman:1973sx}. We reproduce here the line of reasoning as some of this is needed later. 

To leading order in perturbation theory,  a scalar theory with quartic interactions  \eqref{Yukpot} has the beta function \cite{Cheng:1973nv}
\begin{align}\label{lambda}
	\beta_{ABCD} &= \frac{1}{8}\sum_{\{ABCD\}}  \lambda_{ABEF}\ \lambda_{EFCD}\,,
\end{align}
where $\beta_{ABCD}\equiv\mu\partial_\mu\lambda_{ABCD}$ with $\lambda$  fully symmetric in its indices, and the sum running over all permutations. For clarity, in the following we shall write out any index sums explicitly. Vacuum stability requires that for each $A$ we must have $\lambda_{AAAA} \geq 0$, or else the potential becomes unbounded in the $\phi_A$ direction. 
Together with  \eqref{lambda} we have 
\begin{align} \label{eq:scaldiag}
	\beta_{AAAA} \propto 
	\sum_{B,C} 
	\lambda_{AABC}\ \lambda_{AABC} \geq 0\,,
\end{align}
showing that vacuum stability is incompatible with asymptotic freedom, for which we would need this beta function to be negative, \eqref{negative}. 
Let us then switch off all such couplings identically, $\lambda_{AAAA}(\mu) = 0$. In this scenario, their flows and all couplings appearing on the right-hand-side of \eqref{eq:scaldiag} have to vanish, or else a non-zero value for $\lambda_{AAAA}$ is generated by fluctuations. Specifically, taking $B = C$ it follows that $\lambda_{AABB}(\mu) = 0$ at all scales, which again necessitates $\beta_{AABB} = 0$. Since these beta functions are  the sums of squares, 
\begin{align}\label{ABAB}
	\beta_{AABB} &= \beta_{ABAB} \propto 
	\sum_{C,D} 
	\lambda_{ABCD}\ \lambda_{ABCD}\ge 0\,,
\end{align}
the pattern percolates: each and every coupling appearing on the right-hand-side vanishes, $\lambda_{ABCD}(\mu) = 0$,
and the theory remains free at all scales  \cite{Coleman:1973sx}. Thus, we conclude that the Coleman-Gross theorem holds true for  theories with Weyl fermions, and asymptotic freedom cannot be achieved without non-abelian gauge fields.

{\it Price of  interacting fixed points.---}
We are now in a position to discuss the role of gauge field fluctuations for asymptotic safety and weakly coupled fixed points in general, renormalisable theories in four dimensions.
At weak coupling, anomalous dimensions are small and canonical power counting remains applicable. It  is then sufficient to establish weakly-coupled fixed points $\left(g_*,{Y_*},\lambda_*\right)$ for the canonically marginal couplings of the theory, which are the perturbatively-controlled solutions of 
\begin{align}\label{zero}
\mu\,\partial_\mu\left(g,{Y},\lambda\right)|_*= 0\,,
\end{align}
other than the Gaussian,  where at least some or all couplings are non-zero \cite{Bond:2017tbw}. For general gauge theories, a full classification  of  weakly coupled fixed point solutions to \eqref{zero}  has  been given in \cite{Bond:2016dvk}. Perturbative fixed points are either free (the Gaussian), or interacting in the gauge sector (Caswell--Banks-Zaks fixed points) \cite{Caswell:1974gg,Banks:1981nn},  or simultaneously interacting in the gauge and the Yukawa sector (gauge-Yukawa fixed points). Fixed points may be partially or fully interacting, depending on whether some or all gauge couplings take non-zero values. 
Scalar self interactions must take free or interacting fixed points of their own, compatible with vacuum stability (Tab.~\ref{Tab}).
We stress that any weakly-interacting  fixed point 
is controlled by the one-loop gauge coefficient 
 in \eqref{gauge}, irrespective of its sign. Its smallness for suitable matter ensures strict perturbativity \cite{Bond:2016dvk,Bond:2017tbw}. Banks-Zaks fixed points, if they exist,  are always infrared. The Gaussian and gauge-Yukawa fixed points can be infrared or ultraviolet. In particular, asymptotic safety at weak coupling arises solely via gauge-Yukawa fixed points \cite{Bond:2016dvk}.
 We conclude that weakly interacting fixed points  and  asymptotic safety in  non-abelian gauge theories with  matter have the exact same  origin as asymptotic freedom. 

 {\it No-go-theorem for scalar-Yukawa fixed points.---} 
 In order to  complete our  claim, and inasmuch as asymptotic freedom cannot arise  without non-abelian gauge fields, we finally must show that weakly interacting fixed points cannot arise in the absence of gauge interactions. To that end, we return to  scalar-Yukawa theories with interaction Lagrangean \eqref{Yukpot}. Assuming that Yukawa and scalar couplings are small, we must have $\mu\partial_\mu \ybold^A|_* = 0$ at the leading non-trivial order in perturbation theory. Consequently, the bounds \eqref{eq:tracebounds}, \eqref{bound} must be saturated. However,  \eqref{eq:tracebounds} only vanish for vanishing Yukawa couplings,
\begin{align}\label{Y=0}
	\ybold^A_{JK} &= 0\,.
\end{align}
This  is understood as follows. Being a sum of absolute values squared, the expression \eqref{eq:T12} vanishes if and only if each term in the final sum vanishes,
$\ybold_{\!\bm2\,JK}^{\,\,\,\,AB} = 0$.
From the definition for $\ybold_{\!\bm2}$, and after contracting over scalar indices we find that the matrix 
	$\ybold^{\!\bm F}_{\!\bm2}$ also vanishes. Taking its trace 
$ \ybold^{\!\bm F}_{\!\bm2\,JJ} 
	 =\overline{\ybold^A_{JK}}\ybold^A_{JK}=0$
implies \eqref{Y=0} and the vanishing of \eqref{eq:T12} and \eqref{eq:T23}. We conclude that the only available fixed point in the Yukawa sector at one-loop, without gauge fields, is the Gaussian, and it must be infrared.

\begin{table}[t]
\begin{tabular}{ccc}
 \toprule
\rowcolor{Yellow}
\bf \ Case \  &\bf Condition
& \bf \ Fixed Point \ 
\\
\midrule 
 $i)$& $g_i=\ybold^A_{JK}=\lambda_{ABCD}=0$&Gaussian  \\
 $ii)$&   some $g_i\neq 0$, all $\ybold^A_{JK}=0  $&Banks-Zaks \\
 $iii)$& some $g_i\neq 0$, some $\ybold^A_{JK}\neq 0$&gauge-Yukawa\\
\bottomrule
\end{tabular}
\caption{\label{Tab} 
Fixed points of  general weakly interacting quantum field theories in four dimensions.
In cases $ii)$ and $iii)$, scalar self-interactions, if present, must take   fixed points $\lambda^*_{ABCD}$ compatible with vacuum stability \cite{Bond:2016dvk}.}
\end{table} 

Once more, scalar couplings cannot upset this conclusion: scalar selfinteractions contribute to the running of Yukawas starting at two loop. In principle, they could balance the one-loop Yukawa terms provided they are parametrically  larger while still remaining perturbative in their own right. For such a mechanism to be operative, some scalar quartics must take weakly interacting fixed points by themselves. Under this assumption we can  safely neglect the parametrically smaller Yukawa contributions.
Let us then pick $A,B$ such that for some $C,D$ we have $\lambda_{ABCD}^*\neq 0$. This implies  the strict inequality
\begin{align}\label{sum}
	\sum_{C,D} \lambda_{ABCD}^*\ \lambda_{ABCD}^*> 0\,,
\end{align}
as this is a sum of squares of which at least one entry is non-zero.  Combining \eqref{sum} with \eqref{ABAB} we conclude that the flows of $\lambda_{ABAB}$ and $\lambda_{AABB}$ are strictly positive, which is in conflict with \eqref{zero}, and the assumption of a weakly coupled fixed point in the scalar sector cannot be maintained. 
This establishes that  the sole  perturbatively-controlled  fixed point is the Gaussian, which  is  invariably infrared. Ultimately,  in any scalar-Yukawa theory the unavailability of  weakly interacting fixed points and asymptotic safety has the same  origin as the unavailability of asymptotic freedom. 

{\it Discussion and Outlook.---}
In this Letter, we have investigated free or weakly interacting fixed points of $4d$ particle theories with gauge fields, fermions, or scalars. From the viewpoint of high-energy  physics, our findings establish that asymptotic freedom and asymptotic safety are two sides of one and the same medal.
Quantum fluctuations of matter fields alone, with or without photons, are incapable of generating a well-defined and predictive short-distance limit at weak coupling. Rather, the unique driver for viable ultraviolet completions -- $i.e.$~fixed points of the renormalisation group with asymptotic freedom or asymptotic safety -- are the fluctuations of non-abelian gauge fields.
We emphasize that 
abelian gauge theories 
cannot generate   strictly perturbative fixed points in the way non-abelian theories can.
Still, abelian factors
may change from infrared free to asymptotically free
 in the vicinity of partially interacting fixed points, very much like  infrared free {non-abelian} gauge factors in semi-simple theories  \cite{Bond:2016dvk,Bond:2017lnq,Bond:2017suy} (see also \cite{Bond:2019}).  We conclude that non-abelian gauge fields are the price for {\it any} particle theory to remain  strictly perturbative and predictive at asymptotically high energies. 
 We also note that abelian and non-abelian 
 gauge theories with matter may  achieve fixed points at moderate 
 coupling, 
 offering new    directions
  for Standard Model  extensions    \cite{Bond:2017wut,Kowalska:2017fzw}.

From the viewpoint of 
statistical physics, our results show that weakly-coupled infrared fixed points and second order quantum phase transitions  cannot arise without gauge fields. It follows that the universality class for any such phase transition must contain non-abelian gauge interactions as a source for conformality.  
This covers   
conventional  Landau-Ginsburg-type  phase transitions 
 with well-defined order parameters \cite{ZinnJustin:2002ru},
 conformal phase transitions of the Kosterlitz-Thouless type or in the vicinity of fixed point mergers \cite{Miransky:1996pd},
or topological  ones \cite{Senthil:2004aza}
with deconfined quantum critical points 
whose existence
 in four dimensions, once more, 
relates to free or  interacting gauge fields  \cite{Bi:2018xvr}. 
Most notably,  the  findings of this work have shown that the set of necessary and sufficient conditions for weakly interacting fixed points, as  stated in    \cite{Bond:2016dvk}, is complete, 
opening a door for a systematic classification of  critical four-dimensional theories

In a related vein, our results  have also  implications for conformal field theories (CFTs) in four dimensions. Here, conditions under which scale invariance  
entails full conformal invariance  are of particular interest \cite{Polchinski:1987dy,
Komargodski:2011vj,Komargodski:2011xv,Luty:2012ww}.
 Using techniques related to the proof of the $a$-theorem \cite{Komargodski:2011vj,Komargodski:2011xv},  it has been demonstrated that any relativistic and unitary four-dimensional theory that remains perturbative in the ultraviolet or infrared asymptotes to a CFT \cite{Luty:2012ww}. Since all weakly interacting fixed points discussed in this work
(Tab.~\ref{Tab}) belong to this category \cite{Bond:2016dvk} we are lead to the important conclusion that elementary non-abelian gauge fields 
are  the price for interacting, unitary, and strictly perturbative CFTs in four dimensions.
Moreover, the precise quantitative  link between fixed points of the  renormalisation group  and CFTs \cite{cardy_1996}   
can now be used 
to extract
  conformal data 	including  
    scaling dimensions \cite{Litim:2014uca,Bond:2017tbw,Bond:2017lnq,Bond:2017suy} and  structure coefficients   
  \cite{Codello:2017hhh}.
 This   is 
 complementary to the conformal bootstrap approach, which
 exploits  representations of the conformal algebra 
and their short-distance behaviour  \cite{Poland:2018epd}
without being  sensitive to 
the presence or absence
of 
gauge fields in the microscopic theory. 
We conclude that 
our results offer a direct  route to identify and characterize 
many new CFTs in four dimensions from first principles.

It would be most useful to 
also 
clarify the availability
interacting fixed points and  asymptotic safety in  more strongly-coupled $4d$ theories, both with and without quantised gravity 
 \cite{Reuter:1996cp, Litim:2003vp,Niedermaier:2006ns,Litim:2011cp, Niedermaier:2010zz,Benedetti:2009gn,Falls:2013bv,Falls:2014tra,Falls:2016wsa,Falls:2017lst,Percacci:2017fkn,Robinson:2005fj,Shaposhnikov:2009pv,
Folkerts:2011jz,Dona:2013qba,Christiansen:2017cxa,Eichhorn:2018yfc}.
 This  task, however, is much more  challenging. First and foremost,  
the applicability of canonical power counting can no longer be taken for granted  
and  
non-perturbative tools such as  functional renormalisation, lattice simulations,
supersymmetry, or other, become a necessity. Still,  signatures of near-Gaussian scaling dimensions  in asymptotically safe quantum gravity \cite{Falls:2013bv,Falls:2014tra,Falls:2016wsa,Falls:2017lst} on one side, and 
powerful weak-strong
dualities such as in supersymmetric gauge theories \cite{Seiberg:1994pq}  on the other, 
suggest that more advances
can be made
in the future.

{\it Acknowledgements.---} Parts of this work have been supported by a studentship from the Science and Technology Research Council (STFC).

\newpage

\bibliographystyle{apsrev4-1}
\bibliography{PriceASbib}

\end{document}